\documentclass[aps,nobibnotes,prb,superscriptaddress,twocolumn,floatfix,showpacs,amsmath,amssymb, amsfonts,mathtools]{revtex4-1}
\usepackage{graphicx}
\usepackage[pdftex]{hyperref}
\usepackage{subcaption}
\usepackage{floatrow}
\usepackage{latexsym}
\usepackage{psfrag}
\usepackage[makeroom]{cancel}
\usepackage{dcolumn}
\usepackage{blindtext}
\usepackage{epstopdf}
\floatstyle{plaintop}
\restylefloat{table}

\usepackage{amsmath,latexsym,psfrag}
\usepackage{array}
\usepackage{dcolumn}
\usepackage{bm}
\usepackage{soul}  
\usepackage{gensymb}

\begin{document}

\title{Ab Initio Study of K${_3}$Cu${_3}$P${_2}$ Material for Photovoltaic Applications}
\author{Mwende Mbilo}
\affiliation{Department of Physics,Faculty of Science and Technology,University of Nairobi,P.O. Box 30197 - 00100, Nairobi,Kenya}
\email[Corresponding Author:~]{mwendebilo@students.uonbi.ac.ke}
\author{George S. Manyali}
	\affiliation{
		Computational and Theoretical Physics Group (CTheP), Department of Physical Sciences,
		Kaimosi Friends University College, P.O Box 385-50309, Kaimosi, Kenya.}
\author{Robinson J. Musembi}
\affiliation{Department of Physics,Faculty of Science and Technology,University of Nairobi,P.O. Box 30197 - 00100, Nairobi,Kenya}
	\date{\today}
\begin{abstract}
Search for efficient materials for application in the fields of optoelectronics and photovoltaics are active areas of research across the world. The potential of compounds such as K${_3}$Cu${_3}$P${_2}$ is not yet fully realized. Therefore, we perform the ab initio studies based on density functional theory to investigate the structural, electronic, elastic, and optical properties of K${_3}$Cu${_3}$P${_2}$. Ground state properties were computed in three different scenarios, i.e: with spin-orbit coupling (SOC), without spin-orbit coupling, and with Hubbard U parameter. Direct electronic bandgaps of 1.338~eV, 1.323~eV and 1.673~eV were obtained for K${_3}$Cu${_3}$P${_2}$ without SOC, K${_3}$Cu${_3}$P${_2}$ with SOC and K${_3}$Cu${_3}$P${_2}$ with Hubbard U respectively. In all the cases, Cu-d orbitals were dominant at the top of the valence band. The effect of SOC on the K${_3}$Cu${_3}$P${_2}$ computed lattice constant and bandgap was insignificant. The mechanical stability test indicated that K${_3}$Cu${_3}$P${_2}$ is mechanically stable at zero pressure. The optical band gap was found to increase by 0.635~eV when Hubbard U was taken into consideration. Generally, the inclusion of the Hubbard U parameter in density functional theory improves the predictions of the bandgap and optical properties.
\setlength{\parskip}{1em}
\setlength{\parindent}{0pt}

\setlength{\parskip}{1em}
\setlength{\parindent}{0pt}

\end{abstract}
\maketitle

\section{Introduction}
Interest in ternary semiconductor compounds is on the rise due to their wide range of photovoltaic applications \cite{1}. Currently, the semiconductor market is dominated by traditional inorganic materials comprising of Silicon, Germanium, type III-IV compounds and Gallium Arsenide among others \cite{2}. These traditional semiconductors have been studied widely and great milestones in terms of performance properties when incorporated to photovoltaic devices have been achieved \cite{3}. Due to the rise in global energy demand, the evolution of new semiconductor absorber materials for energy devices is increasing. New class of materials including  chalcogenides, oxides, pnictides, and mixed-anion compounds have emerged proving their outstanding performance as compared to their traditional counterparts \cite{4}.  
\setlength{\parskip}{1em}
\setlength{\parindent}{0pt}

Among these materials, pnictides have been less studied with the least power conversion efficiency of 3.4$\%$ reported for ZnSnP$_{2}$ \cite{5}. A study was carried out on ZnXPn$_{2}$ (X = Si, Sn Ge and Pn = P, As) materials \cite{6}. This study showed that pnictides can be applied as thermoelectric materials. An investigation of thermoelectric properties of ABX$_{2}$ compounds using high throughput screening method has been carried out previously \cite{7}. High values of figure of merit predicted from this study were useful for thermoelectric applications. Another study on thermodynamic properties of indium pnictides was carried out using first principles \cite{8}. The thermodynamic properties including specific heat, vibrational free energy, internal energy as well as entropy revealed pnictides as good materials for optoelectronics. Recently, first principles study on properties of K${_3}$Cu${_3}$P${_2}$ and  K$_{3}$Ni$_{3}$P$_{2}$ pnictides was carried out \cite{9}. The calculated properties from this study revealed the potentiality of these materials for thermoelectric as well as optoelectronic applications. Despite these findings, not much is known about the effect of using different DFT methods on pnictides properties. In this study, we carried out structural, electronic, elastic, and optical properties calculation of  K${_3}$Cu${_3}$P${_2}$ using ultrasoft pseudopotentials within generalized gradient approximation (GGA) and GGA + U and compared the results. This work is arranged as follows; section 2 gives the description of the computational methods used, section 3 discusses the major findings and section 4 gives a summary of the computed structural, electronic, elastic, and optical properties.
\section{Computational Methods}
In this study, all the calculations were performed using DFT \cite{28,29} method as implemented in Quantum Espresso (QE) package \cite{10}. Perdew–Burke–Ernzerhof (PBE) functionals were used in this study \cite{11}. The Hubbard U model was employed to approximate the exchange correlation potential in Kohn-Sham equations \cite{12}. The DFT+U calculations were performed on highly correlated copper element whose Hubbard U parameter was obtained from the literature \cite{13}. K${_3}$Cu${_3}$P${_2}$ has a trigonal crystal structure with lattice parameters a = b = c = 14.0403 a.u; $\alpha$ = $\beta$ = $\gamma$ = $45.322^{o}$ as obtained from materials project \cite{27}. The QE input generator was used to generate the PWscf input file of the K${_3}$Cu${_3}$P${_2}$ crystal structure \cite{14}. The convergence of total energy with cutoff energy, lattice constants as well as k-point sampling was performed to obtain optimum values for further calculations. The variable cell relaxation was performed to optimize atomic positions as well as lattice constants. The optimized properties were used for calculation of ground state structural, electronic, elastic, and optical properties.
\section{Results and Discussion}
\subsection{Convergence Tests}
The convergence tests were performed on K${_3}$Cu${_3}$P${_2}$ in order to obtain accurate values for optimal performance. The following plot shows the convergence of total energy with the cutoff energy and K-points.
\begin{figure}[H]
    \centering
    \includegraphics[width=7cm]{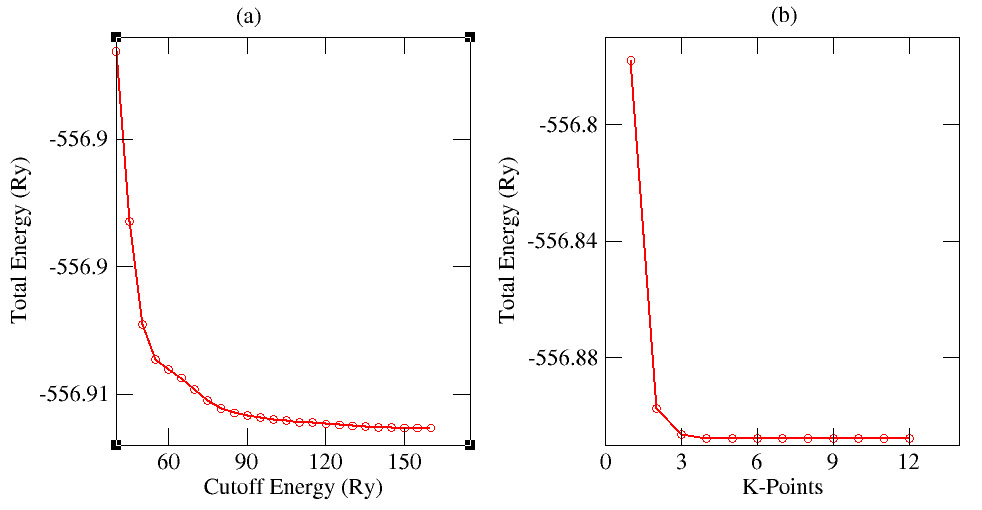}
    \caption{Convergence of total energy with (a) cutoff energy and (b) k-points}
    \label{fig:1}
\end{figure}
From Figure \ref{fig:1}, the total energy converged with cutoff energy at 140 Ry and K-points at 7 x 7 x 7 mesh. The converged values of cutoff energy and k-point mesh were used in the convergence of lattice constants. The output is shown in Figure 2.
\begin{figure}[H]
\centering
    \includegraphics[width=6cm]{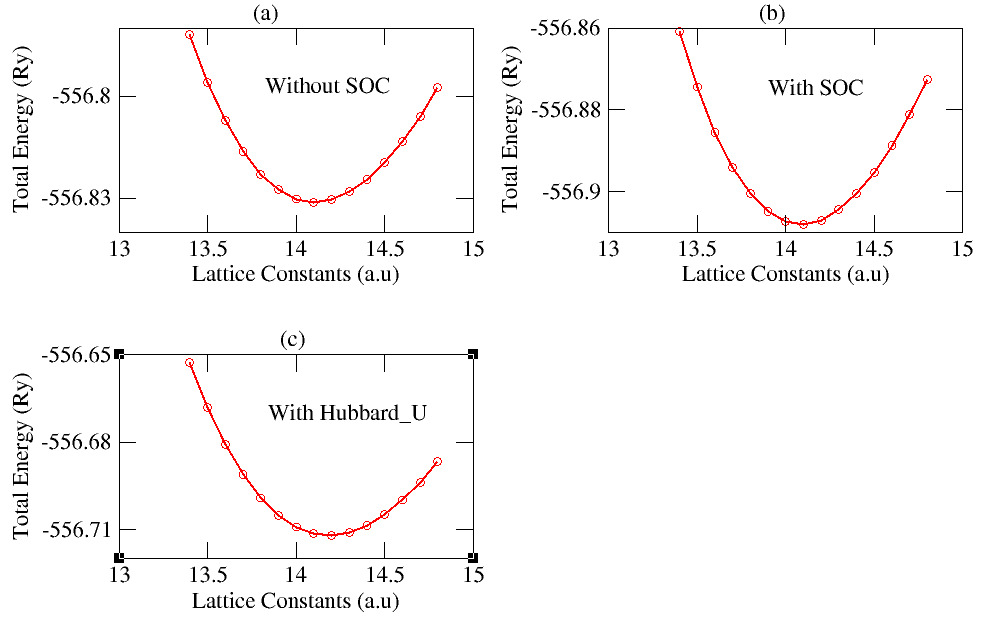}
    \caption{Convergence of total energy with the lattice constants of; (a) K${_3}$Cu${_3}$P${_2}$ without SOC, (b) K${_3}$Cu${_3}$P${_2}$ with SOC and (c) K${_3}$Cu${_3}$P${_2}$ with Hubbard U}
    \label{fig:2}
\end{figure}
The total energy and the lattice constant values were fitted in Murnaghan equation of state \cite{30} in order to determine the structural properties shown in table 1 below.
\begin{table}[H]
\caption{The calculated ground state lattice constants a$_o$(a.u), bulk modulus B$_o$(GPa) and first pressure derivatives B$_o^I$(GPa) of K${_3}$Cu${_3}$P${_2}$ without SOC, K${_3}$Cu${_3}$P${_2}$ with SOC and K${_3}$Cu${_3}$P${_2}$ with Hubbard U}
\begin{tabular}{cccc}
\hline
K${_3}$Cu${_3}$P${_2}$ properties & Without SOC & With SOC &  With Hubbard U\\
\hline
a$_{o}$&14.1043&14.0888&14.1862 \\
B$_{o}$&19.0&19.2&17.8 \\
B$_{o}^I$&0.4566  &0.4500   &0.4516 \\
\hline
\end{tabular}
\end{table}

\subsection{Structure Optimization}
\begin{figure}[H]
\centering
    \includegraphics[width=6cm]{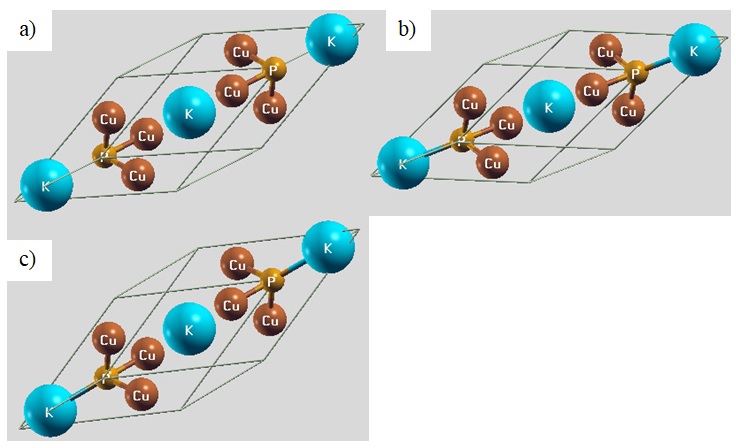}
    \caption{Relaxed trigonal crystal structures; (a) K${_3}$Cu${_3}$P${_2}$ without SOC, (b) K${_3}$Cu${_3}$P${_2}$ with SOC and (c) K${_3}$Cu${_3}$P${_2}$ with Hubbard U}
    \label{fig:3}
\end{figure}

Variable cell relaxation (vc-relax) was performed on K${_3}$Cu${_3}$P${_2}$ without SOC, K${_3}$Cu${_3}$P${_2}$ with SOC and K${_3}$Cu${_3}$P${_2}$ with Hubbard U parameter. The relaxed structures are shown in Figure \ref{fig:3}.
The variation of bond lengths and bond angles of the relaxed structures are shown in the table below.
\begin{table}[H]
\caption{Relaxed bond lengths and angles of (a) K${_3}$Cu${_3}$P${_2}$ without  SOC, (b) K${_3}$Cu${_3}$P${_2}$ with SOC and (c) K${_3}$Cu${_3}$P${_2}$ with Hubbard U}
\begin{tabular}{ c c c c c }
 \hline
 \multicolumn{5}{c}{K${_3}$Cu${_3}$P${_2}$ Material Bonding} \\
\hline
Properties&Before vc-relax & (a)& (b)& (c) \\
\hline
Bond Lengths (a.u) & & & &\\ 
K – P   &6.2059   &6.2665   &6.1987   &6.2087 \\
P – Cu	&4.2627   &4.2710   &4.2675   &4.2989 \\
Bond Angles ($^o$)& & & & \\
K – P – Cu   &132.8   &132.4   &132.9   &132.1 \\
Cu – K – P   &18.9   &19.0   &18.9   &19.3 \\
P – Cu - K	 &28.1   &28.5   &28.1   &28.5 \\
\hline
\end{tabular}
\end{table}
The bond lengths after relaxation were elongated for K${_3}$Cu${_3}$P${_2}$ without SOC and K${_3}$Cu${_3}$P${_2}$ with Hubbard U parameter and reduced for K${_3}$Cu${_3}$P${_2}$ with SOC as compared to bond lengths before relaxation. Similarly, there was a decrease/increase in bond angles after structure relaxation. This shows that relaxation imposed structural changes on K${_3}$Cu${_3}$P${_2}$ crystal structure.
\subsection{Electronic Properties}
The optimized parameters obtained from convergence tests as well as variable cell relaxations were used to calculate the band structure and projected density of states (PDOS). The obtained band structure and projected density of states plots for  K${_3}$Cu${_3}$P${_2}$ without SOC are shown below.
\begin{figure}[H]
    \includegraphics[width=8cm]{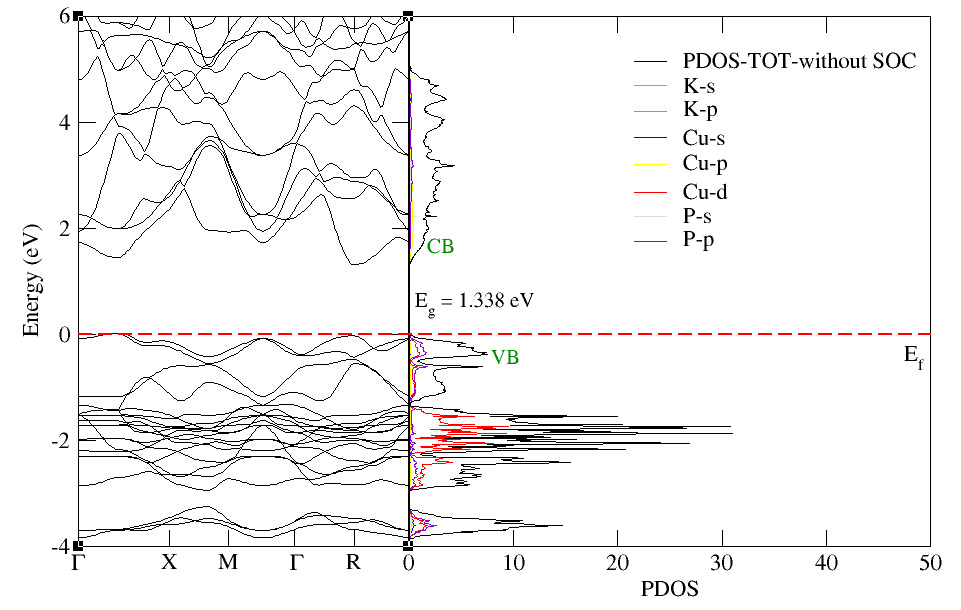}
    \caption{Band Structure and PDOS of K${_3}$Cu${_3}$P${_2}$} without SOC
    \label{fig:4}
\end{figure}
From Figure \ref{fig:4}, the material has a direct bandgap of 1.338~eV. The Cu-d orbitals were dominant followed by P-p orbitals with little contribution in the valence band. The contribution to electronic transitions in the conduction band was by Cu-p orbitals. Hybridization of the other orbitals showed insignificant contribution to both valence and conduction bands. In the presence of the SOC factor, the band structure and PDOS were calculated and plotted as shown in Figure \ref{fig:5}.
\begin{figure}[H]
    \includegraphics[width=7cm]{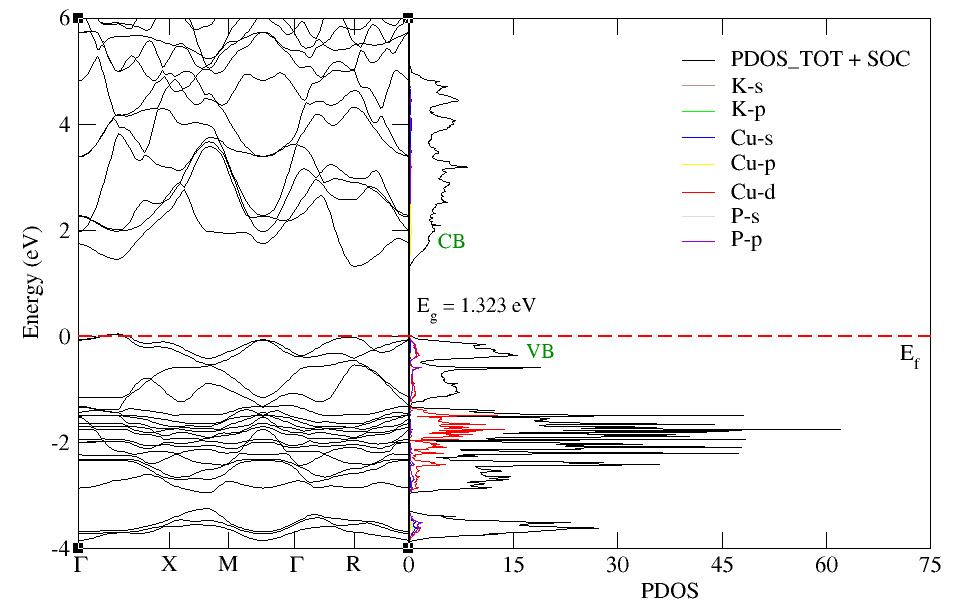}
    \caption{Band Structure and PDOS of K${_3}$Cu${_3}$P${_2}$ with SOC}
    \label{fig:5}
\end{figure}
A direct bandgap of 1.323~eV was obtained for K${_3}$Cu${_3}$P${_2}$ with SOC. This value is slightly lower than the value obtained for K${_3}$Cu${_3}$P${_2}$ without SOC. The contribution to the total electron states was majorly by Cu-d and P-p orbitals in the valence band while Cu-p orbitals contributed to the transitions in the conduction band. Hybridization of other orbitals did not contribute significantly to both the valence and conduction bands. The band structure calculation was also done on the material in presence of Hubbard U parameter. The calculated band structure and PDOS are as shown below.
\begin{figure}[H]
    \includegraphics[width=7cm]{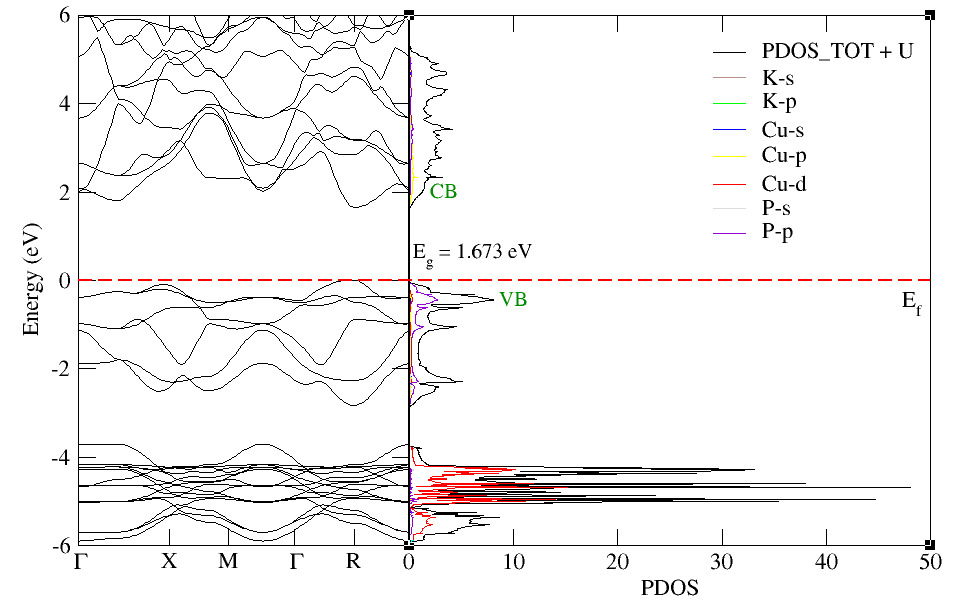}
    \caption{Band Structure and PDOS of K${_3}$Cu${_3}$P${_2}$ with Hubbard U}
    \label{fig:6}
\end{figure}
In presence of the Hubbard U, a direct bandgap of 1.673 eV was obtained. This value is higher than the value obtained for K${_3}$Cu${_3}$P${_2}$ without SOC and K${_3}$Cu${_3}$P${_2}$ with SOC. This is because the Hubbard U parameter introduces correction to the underestimated GGA values to more realistic values obtained experimentally \cite{15}. The bandgap obtained here is slightly below the one reported in the literature \cite{9}. Contribution to the valence band was both by Cu-d and P-p orbitals while contribution to the conduction band was by Cu-p orbitals. The other orbitals contributed insignificantly to both valence and conduction bands. The effect of Hubbard U parameter on bandgap was determined by varying the Hubbard U values as shown below.
\begin{figure}[H]
    \includegraphics[width=7cm]{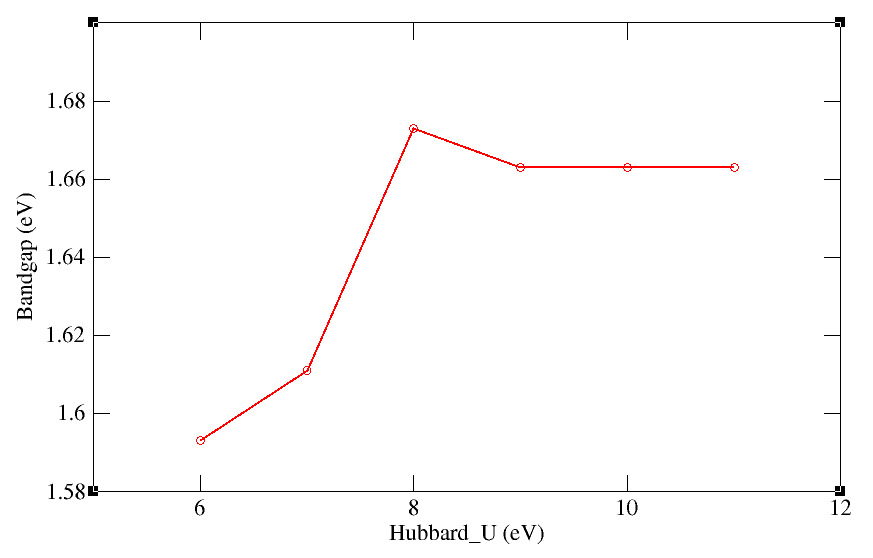}
    \caption{Effect of varying Hubbard U on K${_3}$Cu${_3}$P${_2}$ bandgaps}
    \label{fig:7}
\end{figure}
It is seen in Figure \ref{fig:7} that the optimal Hubbard U value is 8 eV as given in the literature \cite{13}. The range of the values of bandgap obtained are within the range of optimal bandgap values desired for photovoltaic applications \cite{16}.
\subsection{Elastic Properties}
K${_3}$Cu${_3}$P${_2}$ adopts the rhombohedral (I) class (Laue class $\overline{3}m$) crystal structure which features six  independent elastic constants\cite{18}. And the four necessary and sufficient conditions \cite{17,18} for elastic stability in the rhombohedral structure are: 
$C_{11}>|C_{12}|;\\
C_{44}>0; \\
C_{13}^{2}<{\frac{1}{2}}C_{33}(C_{11}+C_{12});\\ 
C_{14}^2<{\frac{1}{2}}C_{44}(C_{11}-C_{12})=C_{44}C_{66}.$
\\From the values of elastic constants calculated in table \ref{table:3}, all the necessary conditions for elastic stability were attained. This confirms that the K${_3}$Cu${_3}$P${_2}$ rhombohedral phase is mechanically stable. To the best of our knowledge; there are no previous works done on the elastic properties of K${_3}$Cu${_3}$P${_2}$ to compare our findings with.  
\begin{table}[H]
\caption{Computed elastic constants C$_{ij}$ (GPa, Voigt-Reuss-Hill Approximations of bulk modulus B (GPa), young's modulus E (GPa), shear modulus G (GPa), Pugh's ratio B/G and Poisson's ratio n for K${_3}$Cu${_3}$P${_2}$}
\begin{tabular}{cc}
\hline
$C_{11}$&37.9\\
$C_{12}$&12.1\\
$C_{13}$&21.3\\
$C_{14}$&9.8\\
$C_{33}$&50.0\\
$C_{44}$&13.9\\
B&25\\
E&25\\
G&9\\
\( \frac{B}{G} \)&2.59\\
n&0.31 \\
 \hline
\end{tabular}
\label{table:3}
\end{table}
The elastic behaviour of materials including ductility and brittleness have been described previously \cite{19}. A Pugh's ratio value of 1.75 determine the ductility/brittleness behaviour. If B/G is less than 1.75, the material is said to be brittle while if B/G is greater than 1.75 the material is termed as ductile. The calculated value of Pugh's ratio in this study show that the material is ductile. Poisson ratio values within the range of 0.1-0.25 and 0.25-0.5 depict covalent and ionic nature of materials respectively \cite{20}. The K${_3}$Cu${_3}$P${_2}$ is ionic with central forces. 
\subsection{Optical Properties}
It is necessary to investigate the optical properties of a material for its prospects in optoelectronic applications. Optical properties describes the frequency response of various optical constants to the incident photon energy \cite{21}. The optical constants in materials are determined by complex dielectric function \cite{21};
$\epsilon(\omega)= \epsilon_{1}(\omega)+ \epsilon_{2}(\omega)$
Where~ $\epsilon_{1}(\omega)$ and $\epsilon_{2}$($\omega$) are frequency dependent real and imaginary parts of dielectric function respectively. The dielectric constants were computed using Sternheimer Equation \cite{22} and presented in Figure 8 below.
\begin{figure} [H]
\includegraphics[width=6cm]{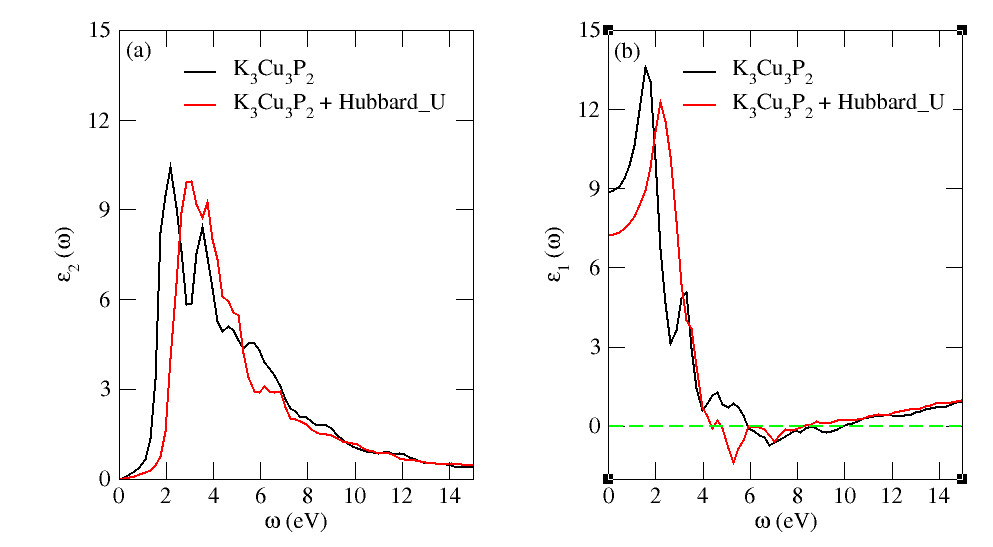}
\caption{The dielectric functions of K${_3}$Cu${_3}$P${_2}$ with and without Hubbard U; (a) Imaginary parts and (b) Real parts}
\label{fig:8}
\end{figure}
The imaginary part of the dielectric function characterizes the photon absorption phenomenon of materials. The peaks in $\epsilon_{2}$($\omega$) plot occur as a result of electron transitions from valence to conduction bands. At $\omega$ = 0 on the y-axis of the $\epsilon_{1}$($\omega$) plot, the values are referred to as static and their square root gives the refractive index values \cite{21}. The optical bandgaps of K${_3}$Cu${_3}$P${_2}$ with and without Hubbard U were found to be 1.839 eV and 1.204 eV while the refractive indices were found to be 2.85 and 3.08 respectively. The narrower the bandgap, the higher the $\epsilon_{2}$($\omega$) and the refractive index values (see Figure 8). These results are in agreement with findings in the previous works \cite{9}. All the other optical constants including the absorption coefficient $\alpha$($\omega$), refractive index n($\omega$), reflectivity R($\omega$) and energy loss function L($\omega$) were obtained from the dielectric constants according to the Equations \cite{23,24,25,26}.
\begin{figure}[H]
\includegraphics[width=7cm]{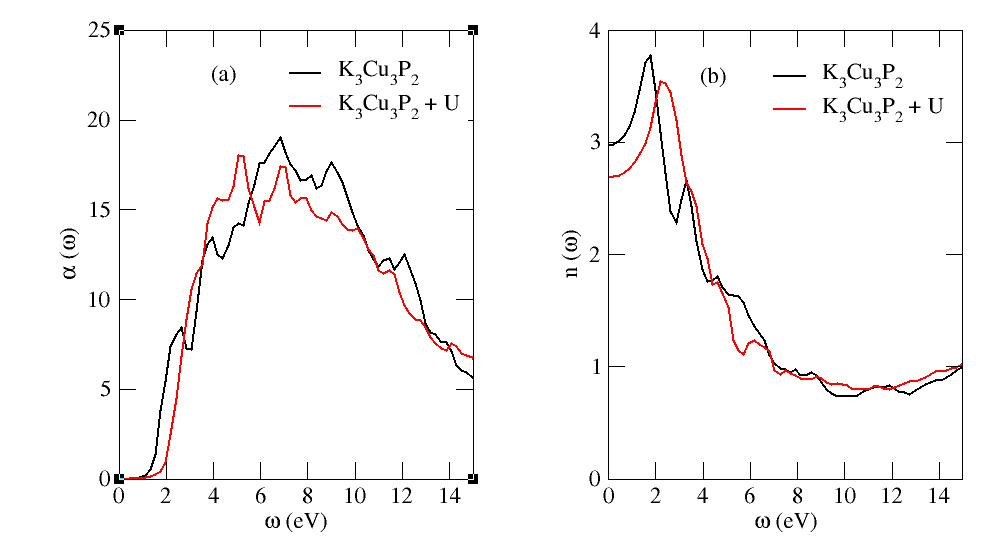}
\caption{(a) Absorption coefficients and (b) Refractive indices of K${_3}$Cu${_3}$P${_2}$ with and without Hubbard U}
\label{fig:9}
\end{figure}
\begin{figure}[H]
\includegraphics[width=6cm]{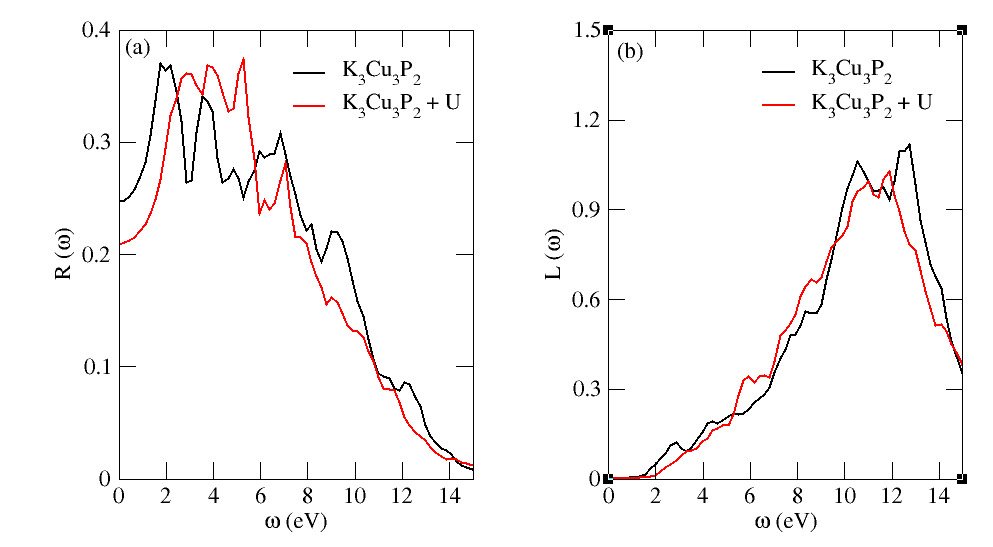}
\caption{(a) Reflectivities and (b) Energy loss functions of  K${_3}$Cu${_3}$P${_2}$ with and without Hubbard U}
\label{fig:10}
\end{figure}
Absorption coefficient measure the quantity of light absorbed by a medium while the refractive index describe both optical and electronic properties of materials \cite{25}. The K${_3}$Cu${_3}$P${_2}$ was found to have a broad absorption spectrum in the range 2-14 eV covering the UV-Vis regions. The major peaks of the refractive index plots were found to lie within the visible region. Reflectivity characterizes the surface behaviour of materials while energy loss spectrum describes the energy lost by fast electrons entering a medium \cite{25}. The main reflectivity peaks were observed at the lower energy regions 2-3 eV. There was a decrease in reflectivity with increase in photon energies. The main peaks of energy loss spectrum were observed at higher energies in the range 10-13 eV. There were no distinct peaks observed in the visible region. The absorption coefficients, refractive indices, reflectivities and energy loss functions of K${_3}$Cu${_3}$P${_2}$ with Hubbard U are lower than those of K${_3}$Cu${_3}$P${_2}$ without Hubbard U (see Figure 9, 10). These findings are in agreement with the results obtained by Irfan et al. \cite{9}.
\section{Conclusion}
In conclusion, we have investigated the effect of using different DFT approaches on the structural, electronic, elastic and optical properties of K${_3}$Cu${_3}$P${_2}$ ternary pnictide material. The material was found to have good structural stability with the ground state lattice constants of 14.104, 14.089 and 14.186 a.u for K${_3}$Cu${_3}$P${_2}$ without SOC, K${_3}$Cu${_3}$P${_2}$ with SOC and K${_3}$Cu${_3}$P${_2}$ with Hubbard U respectively. Direct electronic bandgaps of 1.338 , 1.323 and 1.673 eV were obtained for K${_3}$Cu${_3}$P${_2}$ without SOC, K${_3}$Cu${_3}$P${_2}$ with SOC and K${_3}$Cu${_3}$P${_2}$ with Hubbard U respectively. These results show that SOC factor did not have a significant effect on the computed structural and electronic properties. The contribution to electron transitions in the valence band was majorly by Cu-d orbitals for K${_3}$Cu${_3}$P${_2}$ without SOC, K${_3}$Cu${_3}$P${_2}$ with SOC and K${_3}$Cu${_3}$P${_2}$ with Hubbard U. Optical bandgaps were found to be 1.204 eV and 1.839 eV for K${_3}$Cu${_3}$P${_2}$ without SOC and K${_3}$Cu${_3}$P${_2}$ with Hubbard U respectively. It is noticeable from the optical absorption coefficients that K${_3}$Cu${_3}$P${_2}$ absorbs the photons in the UV-Vis regions of the electromagnetic spectrum hence potentiality for application in the optoelectronic fields.
\section*{Conflict of Interest}
Authors declare no competing conflicts of interest.
\section*{Acknowledgements}
Authors acknowledge the Partnership for Skills in Applied Sciences, Engineering and Technology (PASET) - Regional Scholarship Innovation Fund (RSIF) for the Funding opportunity, the International Programme in Physical Sciences, IPPS Sweden for seed grant for computing resources, Center for High Performance Computing for cluster and software resources, the  Kenya Education Network Trust for funding and Masinde Muliro University of Science and Technology Grant No. MMU/URF/2022/1-026.
\bibliography{main}
\end{document}